\documentclass[reprint, amsmath,amssymb,aps,11pt]{revtex4-1}
\usepackage{csquotes}
\usepackage[utf8]{inputenc}
\usepackage[none]{hyphenat}
\usepackage{physics}
\usepackage{amsmath}
\usepackage{enumerate}
\numberwithin{equation}{section}
\usepackage{graphicx}
\usepackage{bm}
\usepackage[centering,margin=0.8in, includefoot, height=1in, bottom = 1in,]{geometry}
\usepackage{hyperref} 
\usepackage{url}
\hypersetup{
	colorlinks=true,	
	linkcolor=blue,		
	citecolor=red,		
	filecolor=blue,		
	urlcolor=blue		
}
\begin{document}
\def\b{$\bullet $ }
\def\be{\begin{equation}}
\def\ee{\end{equation}}
\pagestyle{plain}
\title{\LARGE Redefining entanglement entropy - can it solve the information paradox?}

\author{\large{Madhur Mehta}}
\email[]{mehta.493@osu.edu}
 \affiliation{Department of Physics, The Ohio State University.}

\begin{abstract}
 Qubit Models which aim to resolve the information paradox must not redefine entanglement entropy as a redefinition does not address the correct issue of the information paradox. The purpose of this letter is to understand why any redefinition of entanglement entropy is not relevant to the paradox. Also, conditions required for a bit model to resolve the information paradox are listed, motivating the community to look for such conditions as a consistency check in their qubit models.
 
\end{abstract}

\maketitle
\section{Introduction}
\noindent
Recently, models which establish the rising and then falling of the Page curve are being considered as solutions to the Information Paradox, and many efforts have been made to show how can the Page curve of a semiclassical black hole be modified to decrease. One such effort is the idea of the \emph{Qubit} Island Entropy \cite{neuenfeld2021double} which provides a bit model to describe the Islands approach for a black hole to bring down its Page curve. The model redefines the Entanglement Entropy for the Hawking radiation which substantially differs from the von Neumann entropy of radiation at later times of evaporation. In this letter, I argue why such a resolution fails to address Hawking's paradox while showing how it describes Hawking's original calculation. Models which redefine entanglement entropy to resolve the information paradox are not relevant to the paradox. In this letter, I show how any redefinition of entanglement entropy could lead to a rising and then falling Page curve, and thus does not solve the paradox.
\\
We begin in section 2, where I present the Hawking's paradox by contrasting the two Page curves produced by a piece of coal and a semiclassical black hole. In section 3, I attempt to understand the Qubit Island Entropy Model in detail and how it differs from the original Hawking's computation. In section 4, I argue how such models do not answer Hawking's paradox. At last, I discuss what conditions must be fulfilled by a correct \emph{qubit} model to do so and end with acknowledgements.
\section{The two Page curves and the Qubit Model}\label{sec2}
\subsection{The paradox}
Hawking's black hole information paradox \cite{hawkingBH} can be cast into the language of Page curves\cite{page1993}. The paradox is the apparent conflict between the Page curve of an ordinary quantum system, say a piece of coal, and the Page curve of a black hole.
In this section, I  discuss the Page curve produced by the burning of a piece of coal and compare it to the Page curve produced by Hawking's computation of a semiclassical black hole.
\subsubsection{Burning Coal}\label{coal}
Let us start with a piece of coal(\emph{N}) which has \emph{n} quantum bits or \emph{qubits}. Let each \emph{qubit} have 2 states, so the total states in the piece of coal becomes $2^n$. 
\begin{figure}[h]
    \centering
    \includegraphics[scale =0.25]{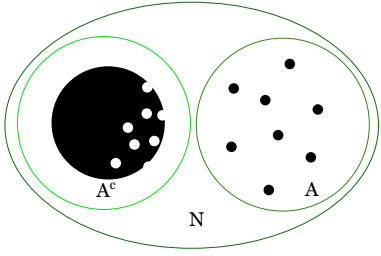}
    \caption{Burning of a piece of coal (system $N$). The set of radiated qubits is labelled as the subsystem $A$ and the rest of the coal is $A^c$}
    \label{fig1}
\end{figure}
\\
We define the qubits emitted out during the burning of the coal, as the subsystem $A$ and we wish to plot the entanglement entropy versus time as the qubits in this subsystem are emitted. 
\\
\\
The von Neumann entropy of the two subsystems $A$ and $A^c$ was calculated in \cite{page1993-1}, \begin{equation}\label{eq1}
    S_A= \log{\mathcal{N}} + O(1/2^m)
\end{equation}
where $\mathcal{N}= \mathrm{min}(N_A,N_{A^c})$ and $N_A =2^m ,N_{A^c}= 2^{n-m}$ are the number of states in $A$ and $A^c$ respectively. The plot of $S_A$ with the \emph{qubits} $m$ gives the Page curve. See Figure \ref{fig2}(a)
\begin{align}
\begin{split}
S_A  \quad= \qquad m \qquad\qquad m\leq n/2
\\
 n-m \qquad\qquad m\geq n/2
\end{split}
\end{align}
\begin{figure}[h]
    \centering
   \includegraphics[scale =0.44]{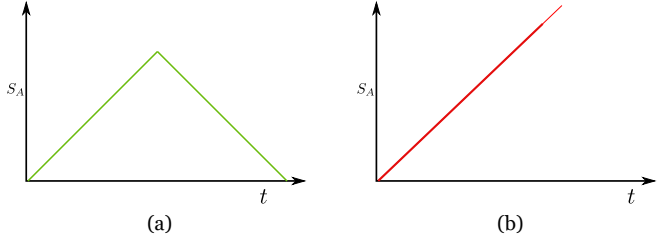}
   \caption{(a)Page curve of a piece of coal which falls down following equation\eqref{eq1}.This is how any quantum system should behave. (b) Page curve of a semiclassical black hole which keeps on rising and does not behave correctly as a quantum system.}
    \label{fig2}
\end{figure}
\subsubsection{Hawking's Black Hole}\label{Hbh}
While the Page curve of any quantum system rises and then falls, the Page curve of a semi-classical black hole differs significantly from this and it is this that leads to Hawking's paradox.
\\
The semiclassical black hole differs from a piece of coal in that it has a horizon where particle pairs are created. The emitted radiation, called Hawking radiation, is entangled with the black hole and as the black hole evaporates, more and more qubits are radiated. The entanglement entropy is a direct measure of the number of entangled qubits, as the black hole evaporates, the entanglement entropy continues to increase. Hence, the Page curve of the semiclassical black hole never falls back to zero(Figure \ref{fig2}(b)).
\\
\\
As opposed to a burning coal, where the radiated qubits carry the information of the internal qubits away at later times, the interior of a black hole is inaccessible and no information can be extracted; therefore, the Page curve never drops. This is one of the forms of the \emph{information paradox}.
\subsection{The Qubit Island Entropy Model(QIEM)}\label{sec4}
\begin{figure}[h]
    \centering
    \includegraphics[scale =0.4]{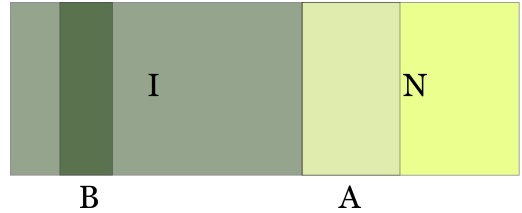}
   \caption{Subsets A of N and B of I are shown. The \emph{Qubit} Island Entropy attains minimum when B is empty($\phi$) for $m\leq n/2$ and when B is all of I for $n/2\leq m \leq n$ where $m$ is the number of bits in A. }
    \label{fig4}
\end{figure}

A resolution of the information paradox in the \emph{qubit} language means a model which brings the Page curve of the semiclassical black hole back to zero as expected from a conventional quantum system. Recent work on \emph{Islands} has tried to show this using replica wormholes \cite{shenker}. The bit model in Ref \cite{neuenfeld2021double} is an attempt to look at Hawking's results and understand the recent Island proposals \cite{islands} which reproduce a rising and then falling Page curve using \emph{qubits} and hence a potential resolution to the information paradox.
\\
In this section, we closely follow Ref\cite{neuenfeld2021double} and rederive Hawking's result using the black hole radiation (name it $N$) along with its interior $I$ and understand the QIEM.
\\
The model and its prescription, as we understand, are as follows:
\begin{enumerate}
    \item Consider the black hole radiation $N$ with $n$ qubits and along with this another system $I$ with $k$ qubits ($k>n$) as in the Figure \ref{fig4};
    \item Mix the systems $N$ and $I$ using some random unitary matrix($U$) and get a system $N\cup I$;
    \item Now, we calculate the entanglement entropy, once using the usual von Neumann method and, once with the QIEM prescription, and compare the two results.
\end{enumerate}
\subsubsection{Paradox in the qubit language}
In Ref \cite{neuenfeld2021double}, Hawking's paradox is described in the following sense. When the two systems are scrambled, we get a system $N\cup I$ with $n+k$ bits. When one looks at a subsystem $A$, it is entangled with its complement $\bar{A}$ which consists of bits from \emph{N} and bits from \emph{I}, and the plot of von Neumann entropy $S^{vN}_A(\tilde{\rho})$ gives the Page curve.
\begin{align}
\begin{split}
S^{vN}_A(\tilde{\rho})  \quad= \qquad m \qquad\qquad m\leq (n+k)/2
\\
 n+k-m \qquad m\geq (n+k)/2
 \end{split}
\end{align}
but since $n<k \rightarrow n<(n+k)/2$ and we see that 
\begin{equation}
S^{vN}_A(\tilde{\rho})  \quad= \qquad m \qquad\qquad m< n
\end{equation}
and the Page curve keeps rising for the first $n$ bits just like the Hawking's paradox as in Figure \ref{fig2}(b). Hence, the addition of the interior resembles Hawking's result.
\subsubsection{Resolution in the qubit language}
Once, Hawking's result has been cast into the \emph{qubit} language, the QIEM computes the quantity which reproduces the falling Page curve:
\be
\tilde{S}_{island}(\tilde{\rho}) = \mathop{min}_{B\subset I} {S^{vN}_{A\cup B}(\tilde{\rho})}
\ee
here $B$ represents an arbitrary set of bits in $I$.
\\
\\
\textbf{To summarize:}
\\
The entanglement entropy of the radiation is redefined using this prescription:
\begin{enumerate}
    \item Choose a system \emph{I} with more bits than the radiation system \emph{N};
    \item Define a small variable part of \emph{I} as \emph{B};
    \item Find the von Neumann entropy of the system $A\cup B$ instead of \emph{A};
    \item Then find the QIEM entropy by choosing that \emph{B} which minimizes the above von Neumann entropy
\end{enumerate}
This gives us 
\begin{align}
\begin{split}
\tilde{S}_{island}(\tilde{\rho})  \quad= \qquad m \qquad\qquad m< n/2
\\
 n-m \qquad n/2\leq m\leq n
 \end{split}
\end{align}
This prescription is understood as a resolution as its prediction resembles the Page curve of a piece of coal.
\section{How the model fails to address the right question}
The above \emph{qubit} islands entropy model describes a falling Page curve which is very interesting, but a closer look at it reveals an important aspect of the information paradox.
\subsubsection{Observable Replacement - the problem}
The QIEM aims to understand how semi-classical gravity emerges, provided there is a true quantum gravitational picture of a black hole. Reproducing the correct Page curve behavior, rising and returning back to zero is understood as the emergence of semi-classical gravity which does not seem correct. Although the QIEM reproduces the falling Page curve for a black hole, it does not resolve the information paradox.
\\
\\
The QIEM prescription does not reproduce a falling Page curve of the von Neumann entropy, rather, it redefines the entanglement entropy as the \emph{Qubit} Island Entropy which produces a Page curve with the desired behavior. The information paradox asks why the von Neumann entropy of a black hole differs significantly from that of a conventional quantum system. Hence, a change to the definition of the entropy for a black hole does not resolve this paradox.
\\
\\
Let us say, we have two systems $A$(which is the coal in our case) and $B$(which is the black hole) and we wish to measure quantity $S_1$(entropy for our case) for $A$ and $B$. Suppose the graphs for $S_1$ for $A$ and $S_1$ for $B$ differ and we wish to resolve this difference. The resolution of such a difference cannot be done by redefining $S_1$ with $S_2$ for system $B$.
\\
 If otherwise, one could just define another type of entropy which would produce a new curve different from both a quantum system and a semiclassical system and call it new Physics for that.
 \\
 So far we have understood that a redefinition of entanglement entropy can resemble a resolution to the information paradox. And now we would see how such a redefinition is not unique and hence cannot be a resolution to the information paradox.
 \subsubsection{Different definitions of Entanglement Entropy}
 Consider again the model as defined in Section \ref{sec4} with two systems namely the radiated qubits \emph{N} and the interior qubits \emph{I}. Let us now look at what another redefinition of the entanglement entropy looks like:
 \\
 \\
 Define: 
\begin{align}
\begin{split}
S_{new}(\tilde{\rho}) = \mathop{min}_{B\subset I} \frac{({S^{vN}_{A\cup B}(\tilde{\rho})})^2}{S^{vN}_A}
\end{split}
\end{align}

\begin{figure}[h]
    \centering
    \includegraphics[scale =0.8]{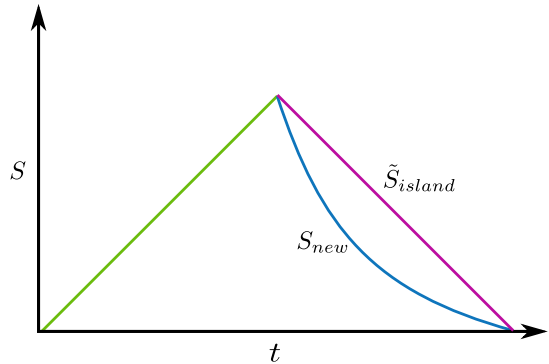}
   \caption{Two kind of Page curves constructed by two different definitions of entanglement entropy. Both the definitions produce a falling Page curve so choosing the correct prescription becomes an issue. A
redefinition of entanglement entropy, thus, cannot
address the information paradox.}
    \label{fig5}
\end{figure}

This definition when applied to find the entropy gives us:
\begin{align}
\begin{split}
S_{new}(\tilde{\rho})  \quad\quad= \qquad m \qquad\qquad m< n/2
\\
 \qquad\quad\frac{(n-m)^2}{m}\qquad  n/2\leq m\leq n,
\end{split}
\end{align}
which gives a rising and falling Page curve but is significantly different from the \emph{Qubit} Island entropy at later times as seen in Figure \ref{fig5}. 
Many such Page curves can be constructed by redefining the entanglement entropy of the radiation and black hole and the above definition is one such example of that. One may equally argue that $S_{new}$ is the correct entanglement entropy as  the Page curve rises at earlier times  and falls at later times. A redefinition of entanglement entropy, thus, cannot address the paradox. One must follow the von Neumann entropy of a quantum system with no change in the definition of entropy while finding a resolution to the paradox through \emph{qubit} models.
\section{Further Discussion}
Hawking's black hole information paradox is the apparent loss of unitary evolution during the evaporation of a black hole. Resolutions to protect unitarity can be created using bit models but they must target the right issues of the paradox.
\\
Recent \emph{Islands} resolutions which aim to tackle the paradox either lead to non-localities  or violate the small corrections theorem as observed in Ref \cite{guo2021contrasting,mathur2009}. The issue of the Page curve of a black hole is the inconsistency between the von Neumann entropy of a burning piece of coal and the von Neumann entropy of a black hole. If this is to be addressed using the \emph{qubit} language, the model must be consistent across both systems (coal and black hole). This leads us to enumerate some requirements which are necessary for a correct \emph{qubit} model which wishes to resolve the information paradox.
\subsubsection*{Requirements for a correct \emph{Qubit} Model to resolve the information paradox}
Any correct model of resolution for the \textit{black hole information paradox} should contain the following :
\begin{enumerate}
    \item The \emph{qubit} model which contains a black hole must have unitary evolution inside the interior of the black hole.
    \item  If the \emph{qubit} model, which consists of a horizon, aims to resolve the paradox, it must show how the effective pair creation at the horizon takes place without violating the effective small corrections theorem \cite{guo2021contrasting}.
    \item The \emph{qubit} model must lead to \enquote{normal}, everyday Physics (classical gravity and thermodynamics) with negligible quantum gravity effects at far away distances from the black hole.
    \item The \emph{qubit} model must generate a Page curve of von Neumann entropy which rises at early times and then falls down to zero after evaporation.
    \item The entanglement entropy should be consistent across all systems and it should not be defined differently for different systems.
\end{enumerate}
The black hole information paradox is a simple yet mysterious problem. Understanding this paradox and working towards a resolution
enables its learner to work in a breeding ground of ideas that connect many different branches of Physics together. The aim of this letter is to motivate the community to understand the correct form of the paradox and create models that tackle the relevant issue.
\\
\begin{acknowledgments}
I wish to thank Prof. Samir D. Mathur for his valuable remarks and motivation. I wish to extend my special thanks to Dominik Neuenfeld and Bin Guo for various discussions and comments. My family which includes Lalit Mehta, Ranu Mehta, Tanvi Mehta and Manvi Mehta was a constant source of motivation. My friends Abhijeet Mishra, Ayan Gupta, Brandon Manley and Naeem Bharmal also helped me to work with dedication and gave important remarks. Support from Kevin Ingles was crucial in making this letter much more comprehensible. Discussions with Marcel Hughes, Sami Rawash, Davide Bufalini, David Turton and Bidisha Chakrabarty were also helpful.
\end{acknowledgments}
\onecolumngrid
\bibliography{letter1}
\begin{center}
    Shanti...
\end{center}
\end{document}